# Gain assisted harmonic generation in near-zero permittivity metamaterials made of plasmonic nanoshells

**Maria Antonietta Vincenti,[1,§,*] Salvatore Campione,[2,§] Domenico de Ceglia,[1,§] Filippo Capolino,[2] and Michael Scalora[3]**

[1]*Aegis Technologies Inc., 410 Jan Davis Dr., Huntsville, AL, 35806, USA*
[2]*Department of Electrical Engineering and Computer Science, University of California Irvine, CA, 92697, USA*
[3]*US Army Charles M. Bowden Research Center, RDECOM, Redstone Arsenal, Huntsville, AL, 35898, USA*
[§]These authors equally contributed to this work
[*]*mvincenti@aegistg.com*

**Abstract.** We investigate enhanced harmonic generation processes in gain-assisted, near-zero permittivity metamaterials composed of spherical plasmonic nanoshells. We report the presence of narrow-band features in transmission, reflection and absorption induced by the presence of an active material inside the core of the nanoshells. The damping-compensation mechanism used to achieve the near-zero effective permittivity condition also induces a significant increase in field localization and strength and, consequently, enhancement of linear absorption. When only metal nonlinearities are considered, second and third harmonic generation efficiencies obtained by probing the structure in the vicinity of the near-zero permittivity condition approach values as high as $10^{-7}$ for irradiance value as low as $10 \text{ MW/cm}^2$. These results clearly demonstrate that a relatively straightforward path now exists to the development of exotic and extreme nonlinear optical phenomena in the $\text{KW/cm}^2$ range.

## 1. Introduction

Finite thickness metamaterials composed of sub-wavelength constitutive elements have stimulated researchers to engineer intriguing homogenized material properties, such as negative refractive index, artificial permeability, and near-zero permittivity (NZP), which in turn have found applications in many practical scenarios. For example, artificial magnetism has been achieved via the split ring resonator (SRR) structure [1], initially introduced at microwave frequencies, and then extrapolated to infrared frequencies by scaling the dimensions of the SRR [2-3]. Plasmonic nanoshells arranged in periodic lattices have been investigated for their unique optical properties, such as strongly directional band gaps [4], negative refractive index [5] and near-zero permittivity [6-8], and various applications, such as Surface Enhanced Raman Scattering (SERS), Surface Enhanced Infrared Absorption (SEIRA) [9] and negative refraction [10]. Moreover, artificial composite materials exhibiting NZP capabilities [11] have attracted the attention of the scientific community in view of their potential uses, including tunneling of electromagnetic energy [12], boosting of optical nonlinearities [13], low-threshold nonlinear effects [14-15], and producing narrow directive beams [16-17].

The main advantage of exploiting the NZP condition to access nonlinear processes is offered by the huge field enhancement that occurs at the interface of a NZP medium. For a lossy NZP medium large field enhancement values due to the continuity requirement of the longitudinal component of the displacement field at the interface are achieved for a TM-polarized incident wave and specific angles of incidence [12, 14]. Recently it has been demonstrated that efficient second (SH) and third harmonic (TH) generation may be achieved in proximity of the zero crossing point of the real part of the dielectric permittivity without resorting to any resonant mechanism [15]. Although Ref. [15] shows several solutions to achieve singularity-driven nonlinear processes in natural materials, losses represent a significant impediment to the triggering of low-threshold, nonlinear phenomena. An efficient way to abate damping in composite, artificial nanostructures consists in the inclusion of active media, such as dyes or quantum dots [6, 8, 18]. For example, in [6] a relative permittivity value of $\text{Im}(\varepsilon) = 10^{-4}$ is estimated in the vicinity of the zero crossing point ($\text{Re}(\varepsilon) = 0$) for an array of gold nanoshells, where silica-like cores including 10 mM of Rhodamine 800 fluorescent dyes pumped at $\lambda \approx 680$ nm provided damping-compensation. Full-wave linear simulations of the metamaterial nanostructure proposed in [6] are shown in Sec.2 as a function of frequency and incident angle of the incoming plane wave. Linear and nonlinear response arising from nanoshells similar to those described in [6] that do not resort to damping compensation have been reported in [19]. The simultaneous availability of the NZP condition and damping-compensation, however, complicates both linear and nonlinear responses of the nanostructure. Here we aim to exploit the enhanced narrow-band processes arising from the metamaterial damping-compensation mechanism to boost the already promising theoretical results reported in reference [15]. We investigate the nonlinear contributions arising from the metal (Sec. 3) and from dielectric media separately, by either filling with, or hosting the nanoparticles in a nonlinear medium (Sec. 4). Although metals do not possess intrinsic, dipolar, quadratic nonlinear terms, they display an effective second order nonlinearity that arises from a combination of symmetry breaking at the surface, magnetic response (Lorentz force), inner-core electrons, convective nonlinear sources and electron gas pressure [20]. We also consider third order nonlinearity of the gold shells that together with effective second order nonlinear sources contribute significantly to the generated signals [21-22]. A detailed study of each nonlinear contribution, without making any a priori assumption about the relative weights of surface and volume sources, allows us to predict SH and TH conversion efficiencies of the order of $10^{-7}$ when probe irradiance is as low as $10 \, \text{MW/cm}^2$. The inclusion of a dielectric nonlinear medium characterized by relatively low second and third order susceptibilities show how to further improve conversion efficiencies paving the way for yet unexplored low-threshold nonlinear phenomena. Finally in Sec. 5 we analyze the critical role of gain-assisted harmonic generation by investigating the effect of different concentrations of the active material. This study serves also as a guide for future experimental works on the subject, where either dye concentration or pump intensity variations can modify significantly the nonlinear response of the metamaterial.

## 2. Linear response of metamaterial slabs composed of plasmonic nanoshells

We start our analysis by considering a composite material having finite thickness $h$ along the $z$ direction, made of four layers of arrayed nanoshells with period $a = 100 \, \text{nm}$, embedded in a homogeneous host medium with relative permittivity $\varepsilon_h = 2.25$ (see Fig. 1). The composite material extends to infinity in the $x$ and $y$ directions, also with period $a$. To mitigate metamaterial losses, the core of the nanoshells is filled with glass doped with Rhodamine 800 fluorescent molecules, homogeneously modeled with relative permittivity according to the relation $\varepsilon_c = 2.25 + \varepsilon_c \left( \Gamma_{\text{pump}}, \overline{N}_0 \right)$, with $\varepsilon_c \left( \Gamma_{\text{pump}}, \overline{N}_0 \right)$ as in [6], assuming the concentration of dye molecules is $\overline{N}_0 = 10 \, \text{mM}$ (other concentration values will be considered in Sec. 5), pumping rate $\Gamma_{\text{pump}} = 6.5 \times 10^9 \, \text{s}^{-1}$, and core radius $r_c = 30 \, \text{nm}$. Each shell is made of gold, with relative permittivity $\varepsilon_s$ modeled through the Drude model as shown in [6], and external radius $r_s = 35 \, \text{nm}$. The relationship between $\varepsilon_c \left( \Gamma_{\text{pump}}, \overline{N}_0 \right)$ and the pump light intensity $I_{\text{pump}}$ is not straightforward when the gain medium is either enclosed by a nanoshell or is close to other nano scatterers (as in the proposed case), whereas it is rather simple for an open gain medium. As a first approximation, one may consider the relation in open gain media where $I_{\text{pump}} = \Gamma_{\text{pump}} h f_{30} / \sigma_{\text{abs}}$, where $\sigma_{\text{abs}}$ is the dye's absorption cross section, $h$ is Planck's constant, and $f_{30}$ is the pump frequency [23]. For Rhodamine 800 dye molecules, $\sigma_{\text{abs}} = 3.14 \times 10^{-16} \, \text{cm}^2$ and $f_{30} = 441 \, \text{THz}$ [24-25], which results in $I_{\text{pump}} = 6.5 \, \text{MW/cm}^2$. Further improvements to the model may be based on a more extensive analysis of the electrodynamic system at the pump frequency and on determining local pump fields and local absorption. We note that the analysis carried out here is performed in the steady state. A more extensive analysis that includes pulsed dynamics and rate equations [23] will be implemented in a future work.

We assume a pump/probe optical setup: the pump, used to activate the dye molecules and set the metamaterial in gain condition, is tuned at $\lambda = 680$ nm (corresponding to 441 THz) and leads to the linear response shown in Fig. 2; we will define the details of the probe signal for the harmonic processes in Sections 3 and 4. More details on the design of the metamaterial under investigation may be found in reference [6]. The main feature of the proposed structure is the achievement of efficient damping-compensation in the vicinity of the zero crossing point of the real part of the metamaterial effective dielectric constant. These two conditions in fact are not coincident in frequency: an effective $\mathrm{Im}(\varepsilon) \approx 10^{-4}$ is obtained for $\lambda \approx 712$ nm while $\mathrm{Re}(\varepsilon) \approx 0$ at $\lambda \approx 710.6$ nm. The effective permittivity has been calculated by means of the Nicolson-Ross-Weir method applied to a metamaterial slab of finite thickness under normal incidence conditions [26-27].

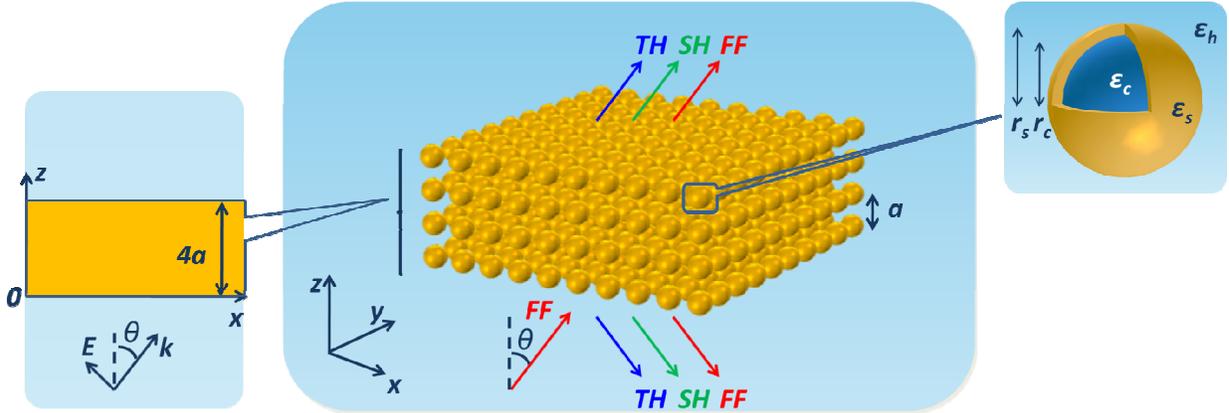

Fig. 1. Sketch of the composite material with finite thickness $h$ made of 4 layers of nanoshells with distance $a$ embedded in a homogeneous medium with relative permittivity $\varepsilon_h$. The composite material extends to infinity in the $x$ and $y$ directions, with period $a$. The nanoshell core is made of glass doped with Rhodamine 800 fluorescent molecules, with homogenized relative permittivity $\varepsilon_c$ and radius $r_c$. The shell is made of gold, with relative permittivity $\varepsilon_s$ and external radius $r_s$. FF stands for fundamental frequency, SH and TH stand for second and third harmonic, respectively.

The metamaterial slab is illuminated with a TM-polarized plane wave, incident at an angle $\theta$ with respect to the $z$ axis, with electric field and wavevector lying on the $x$-$z$ plane (Fig. 1). Transmission, reflection and absorption as functions of wavelength and incident angle are shown in Fig. 2, calculated using two different methods but yielding similar results: a single dipole approximation (SDA) (Fig. 2(a), (c) and (e)) and a full-wave simulation based on a finite-element method in the frequency-domain (COMSOL Multiphysics) (Fig. 2(b), (d) and (f)). The SDA approach [28-30] assumes each nanoshell behaves as an electric dipole, with induced electric dipole moment $\mathbf{p} = \alpha_{ee}\mathbf{E}^{loc}$, with $\alpha_{ee}$ the electric polarizability modeled via Mie theory [28] and $\mathbf{E}^{loc}$ the local field acting on it. The SDA is a good approximation when the plasmonic nanoshells are used close to their fundamental plasmonic frequency, when particle dimensions are much smaller than the incident wavelength, and when the period $a \geq 3r_s$. In general, more accurate results may be obtained by including multipole field contributions [28, 31-33].

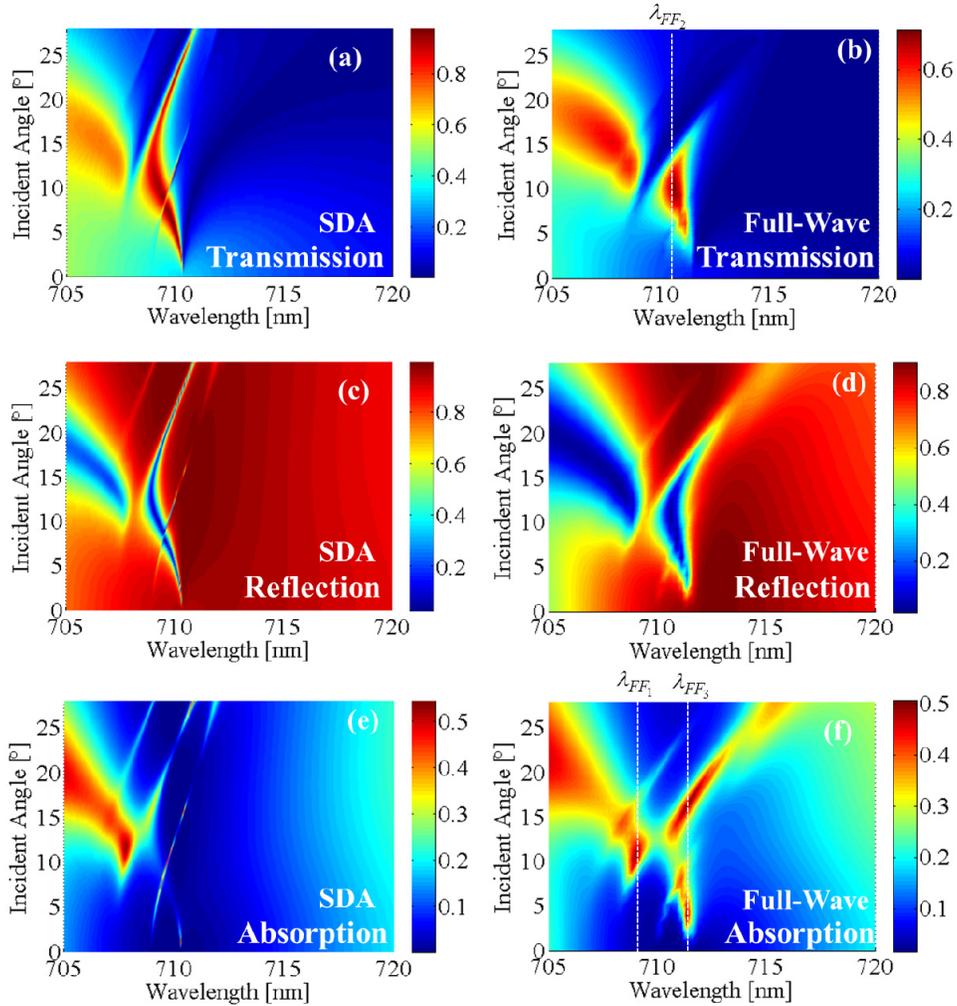

Fig. 2. Transmission ((a) and (b)), reflection ((c) and (d)) and absorption ((e) and (f)) as a function of wavelength and incident angle calculated via SDA approach and full-wave (finite element method) simulation as noted, for the damping-compensated metamaterial slab in Fig. 1.

As shown in Fig. 1, we model $N = 4$ layers of nanoshells embedded in a homogeneous environment. Let us consider a reference nanoshell located at the $n$th layer $\mathbf{r}_n = x_0\hat{\mathbf{x}} + y_0\hat{\mathbf{y}} + z_{0,n}\hat{\mathbf{z}}$ (we can assume, $x_0 = 0$, $y_0 = 0$ and $z_{0,n} = na - a/2$, with $n = 1, ..., 4$). The local electric field acting on it is given by $\mathbf{E}^{\text{loc}}(\mathbf{r}_n) = \mathbf{E}^{\text{inc}}(\mathbf{r}_n) + \sum_{m=1}^{4} \underline{\mathbf{G}}_{n,m}^{\text{layer}} \cdot \mathbf{p}_m$, where $\mathbf{E}^{\text{inc}}(\mathbf{r}) = \mathbf{E}_0 e^{j\mathbf{k} \cdot \mathbf{r}}$ is the TM incident electric field with the wavevector $\mathbf{k}$ lying in the $x$-$z$ plane, and $\mathbf{p}_m$, $m = 1, ..., N$, are the electric dipole moments at positions $\mathbf{r}_m$. Furthermore, the term $\underline{\mathbf{G}}_{n,m}^{\text{layer}} \equiv \underline{\mathbf{G}}^{\infty}(\mathbf{r}_n, \mathbf{r}_m)$, for $n \neq m$, represents the periodic dyadic Green's function (PDGF) for a two dimensional periodic array of dipoles [34-35], whereas $\underline{\mathbf{G}}_{n,n}^{\text{layer}} \equiv \bar{\underline{\mathbf{G}}}^{\infty}(\mathbf{r}_n, \mathbf{r}_n)$ represents its regularized counterpart (without the contribution from the dipole at $\mathbf{r}_n$) [34-35]. Both $\underline{\mathbf{G}}^{\infty}$ and $\bar{\underline{\mathbf{G}}}^{\infty}$ are evaluated through the Ewald method [36-37] that provides: (i) rapid converging summations (only a handful of summation terms is needed to achieve convergence) and (ii) analytic continuation to the complex wavenumber plane. The Ewald method for the two dimensional periodic array scalar Green's functions is detailed in references [38-40]. Following the formulation in Sections III.C and III.D in reference [35], we then compute the induced dipole moments $\mathbf{p}_n$, $n = 1, ..., 4$ by solving the four combined linear equations $\sum_{m=1}^{4} \underline{\mathbf{A}}_{n,m} \cdot \mathbf{p}_m = \alpha_{\text{ee}} \mathbf{E}^{\text{inc}}(\mathbf{r}_n)$, where $\underline{\mathbf{A}}_{n,m} = -\alpha_{\text{ee}} \underline{\mathbf{G}}_{n,m}^{\text{layer}}$, for $n \neq m$, and $\underline{\mathbf{A}}_{n,n} = \underline{\mathbf{I}} - \alpha_{\text{ee}} \underline{\mathbf{G}}_{n,n}^{\text{layer}}$. Once the solution $\mathbf{p}_n$ has been determined, reflection and transmission coefficients are simply evaluated using again the PDGFs $\underline{\mathbf{G}}^{\infty}(\mathbf{r}, \mathbf{r}_n)$. A slight blue shift in the results retrieved via SDA with respect to the full-wave COMSOL result (see Fig. 2) is observed. Also, transmission, reflection and absorption values are slightly different in the SDA case because we are not taking into account the multipolar response. However, similar spectral and angular features are clearly observed in both cases in Fig. 2. The presence of a damping-compensation mechanism, or more generally NZP condition in presence of low losses [41], allows the occurrence of narrow band features and boosts electric field values. We stress that we do not treat the metamaterial as a homogenized slab with effective parameters and that for any incident angle and wavelength the field localization inside the metamaterial slab vary substantially and thus lead to different nonlinear responses.

In order to investigate the more general behavior of the structure in the proximity of the $\text{Re}(\varepsilon) \approx 0$ condition, we calculated the induced dipole moment $\mathbf{p} = p_x \hat{\mathbf{x}} + p_z \hat{\mathbf{z}}$ in the $x$ and $z$ directions at the section $z = 3a/2$, which corresponds to the second bottom layer of nanoshells (Fig. 3). Inspection of Fig. 3 reveals that in the vicinity of the NZP condition, the magnitude of the longitudinal component of the induced dipole moment $p_z$ (Fig. 3 (a)) is much larger than the transverse component $p_x$ (Fig. 3 (b)), which remains small. By looking at the phase of the induced dipole moment in the same spectral/angular region one may also note that the $\mathrm{Re}(\varepsilon) \approx 0$ condition causes a $\pi$-shift in the phase of the longitudinal component, while no phase-shift is observed for the transverse component. These conditions may be important in phase-sensitive processes like harmonic generation and optical bistability.

A key aspect for the purposes of efficient harmonic generation and low-threshold optical bistability processes is also electric field enhancement inside the structure. We used a full-wave approach to calculate the electric field intensity enhancement, defined as $EF = \max\left( |\mathbf{E}|^2 \big/ |\mathbf{E}^{\mathrm{inc}}|^2 \right)$, where $|\mathbf{E}|$ is the total electric field magnitude inside the metamaterial slab and $|\mathbf{E}^{\mathrm{inc}}|$ is the electric field magnitude of the incident plane wave. Our calculations predict the following electric field enhancements: $EF \approx 1000$ at $\lambda_{\mathrm{FF_1}} = 709.5$ nm and $\theta = 13°$; $EF \approx 1900$ at $\lambda_{\mathrm{FF_2}} = 710.6$ nm and $\theta = 7°$; and $EF \approx 3000$ at $\lambda_{\mathrm{FF_3}} = 711.35$ nm and $\theta = 4°$, with angular trends that match well the shape of the absorption profile in the structure (see Fig. 2(f)). Although high field enhancement values are promising for efficient nonlinear processes, one should keep in mind that nonlocal effects may intervene for small metal thicknesses, as may be the case for a 5 nm thick gold shell. However, in what follows we will assume that nonlocal effects are either negligible or may be compensated with higher concentrations of gain material, and postpone this discussion to a future effort.

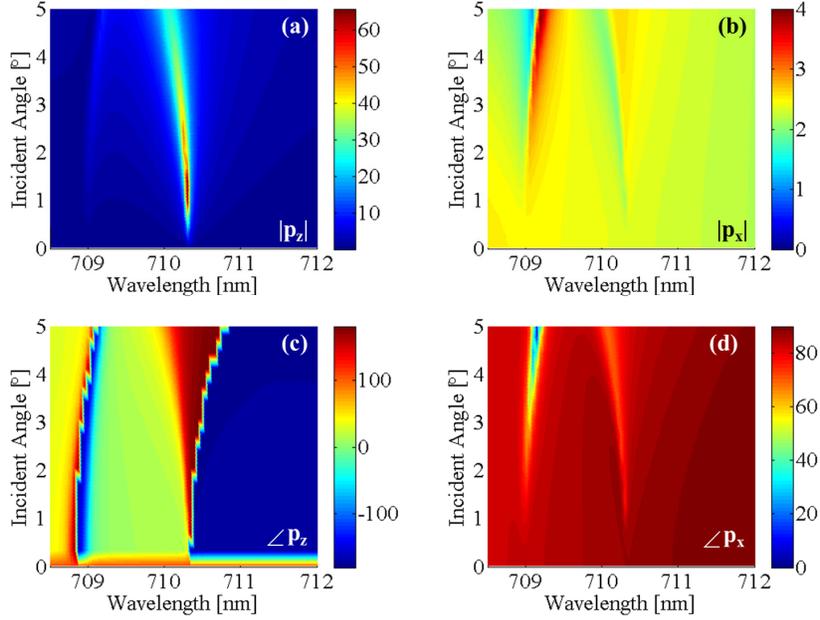

Fig. 3. Induced dipole moment magnitude in the (a) $z$ and (b) $x$ direction at $z = 3a/2$ for the metamaterial slab in Fig. 1. Induced dipole moment phase in the (c) $z$ and (d) $x$ direction at $z = 3a/2$ for the metamaterial slab in Fig. 1. These results have been obtained via SDA approach.

## 3. Nonlinear response of metamaterial slabs arising from metal shells

We now consider a pump-probe scenario, where we assume that a continuous wave (CW) pump signal keeps the system in the damping-compensation regime described in the previous section. We then assume a time harmonic probe signal with irradiance $10 \text{ MW/cm}^2$ and calculate its second and third generated harmonics. This relatively small irradiance guarantees that the probe will remain undepleted so that we may neglect any self-phase-modulation and down-conversion processes in our calculations. Here we also neglect any self-phase modulation of the pump signal and cross-phase modulation between pump and probe signals. Moreover, we expect that harmonic generated signals triggered by the pump field will not significantly alter the harmonic generation conversion efficiencies relative to the probe frequency for two reasons: i) pump and probe are both CW signals so that their harmonics will fall at different frequencies and ii) the pump signal is tuned in a spectral region where the metamaterial absorbs and no substantial field enhancement is expected. For these reasons in what follows we will neglect pump contribution to harmonic generation associated with the probe signal.

### 3.1 Second harmonic generation from gold nanoshells

The nonlinear contributions to second harmonic generation arising from the metal come mostly from free electrons [20, 42-43]. The hydrodynamic model is used to describe free-electron motion, and we do not make any assumptions on the relative weight of surface and volume contributions. A nonlinear equation of motion for the current density of free electrons may be written in the time domain as follows (a tilde (~) is used to denote quantities in the time domain) [20, 42-43]:

$$\frac{\partial \tilde{\mathbf{J}}_f}{\partial t} + \gamma_f \tilde{\mathbf{J}}_f = \varepsilon_0 \omega_p^2 \tilde{\mathbf{E}} + \frac{1}{n_0 e}\Big[\tilde{\mathbf{J}}_f \nabla \cdot \tilde{\mathbf{J}}_f + \big(\tilde{\mathbf{J}}_f \cdot \nabla\big)\tilde{\mathbf{J}}_f\Big] - \frac{\mu_0 e}{m^*}\tilde{\mathbf{J}}_f \times \tilde{\mathbf{H}} \qquad (1)$$

where $\tilde{\mathbf{J}}_f$ is the free electron current density, $\gamma_f = 7.28 \times 10^{13}$ 1/s is the damping coefficient associated with gold losses, $e$ is the electron charge, $m^* = 9.109 \times 10^{-31}$ kg is the effective electron mass, $\mu_0$ is the vacuum magnetic permeability, $n_0 = 4.963 \times 10^{22}$ cm$^{-3}$ is the conduction electron density, $\tilde{\mathbf{H}}$ and $\tilde{\mathbf{E}}$ are the local magnetic and electric fields, respectively.

We transform equation (1) into the frequency domain by expressing all the variables as superposition of two components, one oscillating at the fundamental wavelength and the other oscillating at the SH frequency. We may write the following frequency domain expression for the current density at the SH frequency,

$$\mathbf{J}_{SH} = -i\varepsilon_0\big[\varepsilon_s(\omega_{SH}) - 1\big]\omega_{SH}\mathbf{E}_{SH} + \mathbf{J}_{vol} + \mathbf{J}_{surf}. \qquad (2)$$

The first term on the right hand side of equation (2) is the linear response at the SH angular frequency $\omega_{SH}$ due to both free and bound electrons, and it depends on the experimental permittivity of gold at the SH frequency $\varepsilon_s(\omega_{SH})$. The second ($\mathbf{J}_{vol}$) and third ($\mathbf{J}_{surf}$) terms are bulk and surface nonlinear contributions due to free electrons. These source terms may be derived within the context of the hydrodynamic model in equation (1), following the method illustrated in [44-45] for vacuum-metal interfaces, which is adapted to take into account absorption losses in the metal and the effects of a generic dielectric with permittivity $\varepsilon_B$ in contact with metal as detailed in [46],

$$\mathbf{J}_{vol} = i\omega_{SH}\varepsilon_0 \frac{e}{2m^*\omega_{FF}^2}\frac{\big[\varepsilon_{s,f}(\omega_{FF}) - 1\big]}{2}\beta\Big[\big(\alpha - 1\big)\big(\mathbf{E}_{FF} \cdot \nabla\big)\mathbf{E}_{FF} + \frac{1}{2}\nabla\big(\mathbf{E}_{FF} \cdot \mathbf{E}_{FF}\big)\Big] \qquad (3)$$

$$\hat{\mathbf{t}} \cdot \mathbf{J}_{surf} = -i\omega_{SH}\varepsilon_0 \chi_\parallel^{(2)} E_{FF,\parallel} E_{FF,\perp}$$
$$\hat{\mathbf{n}} \cdot \mathbf{J}_{surf} = -i\omega_{SH}\varepsilon_0 \chi_\perp^{(2)} E_{FF,\perp}^2 \qquad (4)$$

In expressions (3) and (4) $\omega_{FF}$ is the fundamental angular frequency, $\varepsilon_{s,f}(\omega_{FF}) = -\left(e^2 n_0 / m^*\right) \big/ \left(\omega_{FF}^2 + i\omega_{FF}\gamma_f\right)$ is the free-electrons permittivity, $\alpha = \omega_{FF} \big/ \left(\omega_{FF} + i\gamma_f\right)$, $\beta = \omega_{SH} \big/ \left(\omega_{SH} + i\gamma_f\right)$, and $\mathbf{E}_{FF}$ is the electric field at the fundamental frequency; $\hat{\mathbf{t}}$ and $\hat{\mathbf{n}}$ define a local, boundary coordinate system and they represent unit vectors pointing in the directions tangential and outward normal to the metallic surfaces, respectively; $E_{FF,\parallel}$ and $E_{FF,\perp}$ are the magnitude of the tangential and normal electric field components at the fundamental frequency in this local coordinate system. The surface, quadratic nonlinear susceptibilities are then given by

$$\chi_{\parallel}^{(2)} = -\alpha\beta \frac{e}{2m^*\omega_{FF}^2} \left[ \frac{\varepsilon_s\left(\varepsilon_B - 1\right) + \varepsilon_B\left(1 - \varepsilon_{s,f}(\omega_{FF})\right)}{\varepsilon_B} \right]$$

$$\chi_{\perp}^{(2)} = -\alpha\beta \frac{e}{2m^*\omega_{FF}^2} \left[ \frac{3\varepsilon_s^2\left(\varepsilon_B - 1\right) + \varepsilon_B\varepsilon_s\left(\varepsilon_B - \varepsilon_{s,f}(\omega_{FF})\right) + 3\varepsilon_B^2\left(1 - \varepsilon_{s,f}(\omega_{FF})\right)}{4\varepsilon_B^2} \right] \tag{5}$$

where $\varepsilon_B$ respectively assumes value $\varepsilon_c$ or $\varepsilon_h$ depending if the inner or outer interface of the shell is considered for $\chi_{\parallel}^{(2)}$ and $\chi_{\perp}^{(2)}$ calculation.

In Fig.4 we show the forward, backward and total (i.e., forward plus backward) SH conversion efficiencies calculated as $\eta_{\text{SH}} = P_{2\omega} / P_{\omega}$, where $P_{\omega}$ is the input probe power and $P_{2\omega}$ is the radiated power (in the forward, backward or forward plus backward directions) at the second harmonic [47]. We analyzed three different wavelengths where either absorption or transmission peaked ($\lambda_{\text{FF}_1} = 709.5$ nm -Fig. 2(f) -, $\lambda_{\text{FF}_2} = 710.6$ nm - Fig. 2(b) -, $\lambda_{\text{FF}_3} = 711.35$ nm - Fig. 2(f)-). The results in Fig.4(b), (c) and (d) reveal a dramatic increase of the generated signal by at least five orders of magnitude when compared with a flat gold film (Fig.4 (a)) having the same thickness as the shell (5 nm). Moreover, we note that forward and backward harmonic efficiencies are strongly modulated as the incident angle varies. In particular, while one may certainly surmise a relationship between the trends of linear absorption and total SH generation, it is not straightforward to establish a one-to-one relation between the maxima of the forward and backward harmonic processes and absorption or transmission profile. Field localization and phase accumulation at the probe (the probe is tuned in a region where phase accumulation is close to zero) and harmonic frequencies assume completely different character for each peak of transmission/absorption for the probe, so that forward and backward harmonic generation arising from both surface and volume contributions should be examined on a case by case basis and cannot be merely inferred from the linear response of the structure. A rough estimation of the *effective second order nonlinear susceptibility* that comes from symmetry breaking at the surface, magnetic dipoles, inner-core electrons and convective nonlinear sources may be obtained by replacing these terms with a $\chi_{eff}^{(2)} \sim 5000$ pm/V. This value gives the reader an idea of the kind of field enhancement that is achieved inside each gold nanoshell.

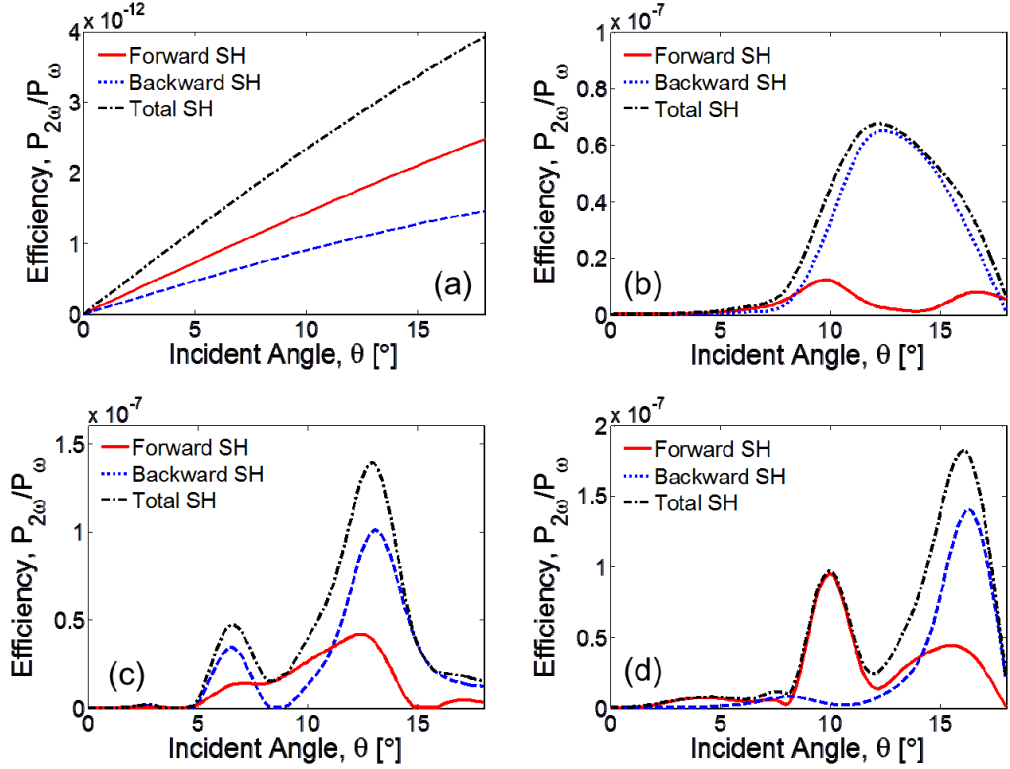

Fig.4. Forward, backward and total (calculated as forward plus backward) second harmonic conversion efficiency $\eta_{SH} = P_{2\omega} / P_{\omega}$ for (a) 5 nm thick flat gold, (b) NZP metamaterial described in Sec. 2 at $\lambda_{FF_1} = 709.5$ nm , (c) at $\lambda_{FF_2} = 710.6$ nm , and (d) at $\lambda_{FF_3} = 711.35$ nm , as a function of incident angle. Note the 5 orders of magnitude increase when using the metamaterial in Sec. 2. These results have been calculated by means of full-wave (finite element method) simulations.

### 3.2 Third harmonic generation from gold nanoshells

Like all other metals, gold possesses a relatively large third order nonlinearity. However, different material configurations, growth techniques and pulse duration have led to a disparate set of values for the third order susceptibility, ranging from $10^{-14}$ m$^2$/V$^2$ to $10^{-19}$ m$^2$/V$^2$ [48-50]. Since we are not specifically interested in quantifying or optimizing conversion efficiency values and instead wish to explore the nonlinear optical properties triggered by the NZP and damping-compensation regimes, in what follows we assume $\chi_{xxxx}^{(3)} = \chi_{yyyy}^{(3)} = \chi_{zzzz}^{(3)} = 10^{-17}$ m$^2$/V$^2$ (all other tensor components are assumed to be zero). As we did for SHG, here too we calculate forward, backward and total TH signal at $\lambda_{FF_1} = 709.5$ nm (Fig. 5 (b)), $\lambda_{FF_2} = 710.6$ nm (Fig. 5 (c)), $\lambda_{FF_3} = 711.35$ nm (Fig. 5 (d)). Conversion efficiency is calculated as $\eta_{TH} = P_{3\omega} / P_{\omega}$, where $P_{3\omega}$ and $P_{\omega}$ are the radiated power at the third harmonic frequency, and the input probe power [47], respectively. A cursory examination of Fig. 5 reveals an increase of the generated third harmonic signal from the nanocomposite metamaterial of about three orders of magnitude when compared with a flat gold film of the same thickness of the shell (5 nm) in Fig. 5 (a). As mentioned above, care should be exercised when relating the features of linear and nonlinear processes. The highly selective angular and spectral features found in both linear and harmonic response may indeed be ascribed to different mechanisms taking place inside the structure. An investigation of the field distribution inside the structure reveals that a number of resonant modes (three in this case) and an impedance matching condition may be identified in this structure, all leading to different degrees of field localization across the slab and consequently different harmonic responses. Moreover, different phase accumulations in the structure due to the NZP regime where the probe is tuned, together with the difference in parity and phases of the excited resonant modes may in turn either maximize or abate field overlap, leading to complicated harmonic angular responses and differences between forward and backward generated signals. Moreover, a different angular dependence of TH signals (Fig. 5) when compared to SH (Fig.4) is also expected based on the different trends of the two nonlinear processes for the 5-nm thick flat gold film (compare Fig.4 (a) with Fig. 5 (a)) and on the fact that they arise either from surface and volume terms or solely from bulk contributions.

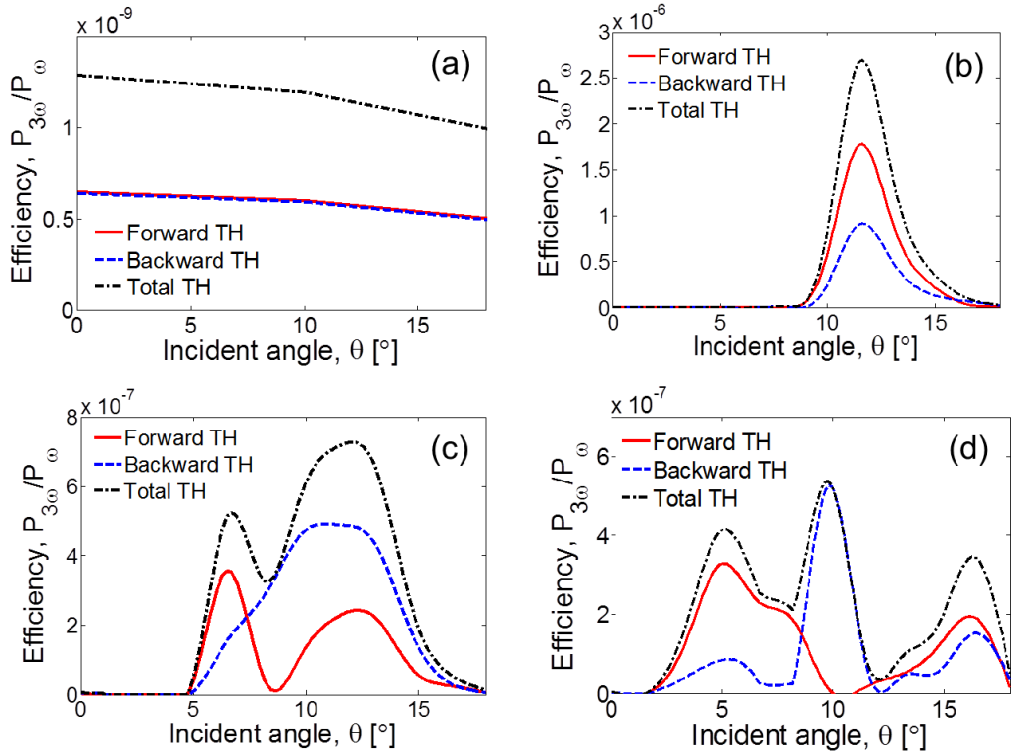

Fig. 5. Forward, backward and total (calculated as forward plus backward) third harmonic conversion efficiency $\eta_{TH} = P_{3\omega} / P_{\omega}$ for (a) 5 nm thick flat gold, (b) NZP metamaterial described in Sec. 2 at $\lambda_{FF_1} = 709.5$ nm, (c) at $\lambda_{FF_2} = 710.6$ nm, and (d) at $\lambda = 711.35$ nm, as a function of incident angle. Note the 3 orders of magnitude increase when using the metamaterial in Sec. 2. These results have been calculated by means of full-wave (finite element method) simulations.

## 4. Nonlinear response of metamaterial slabs arising from bulk nonlinearities

One may now ask the following questions: (i) how does the nonlinear response change if dielectric bulk nonlinearities are also considered in the calculations? and (ii) what is the relative importance of dielectric versus metal nonlinearities, and how much does each contribute to the generated signals? In this section we will attempt to answer these questions by simulating the effect of bulk nonlinearities associated with the dielectric materials in the system, i.e. host medium and shells' core, and by comparing these results with the radiated SH and TH arising only from the gold shells.

To account for nonlinear effects in the host and core dielectrics we express the leading contributions of the nonlinear polarization densities in the $j$ direction at second ($P_{SH,j}$) and third ($P_{TH,j}$) harmonic frequencies as [47]

$$P_{SH,j} = \varepsilon_0 \sum_{l,m=1}^{3} \chi_{jlm}^{(2)}(\omega_{SH}, \omega_{FF}, \omega_{FF}) E_{FF,l} E_{FF,m}$$

$$P_{TH,j} = \varepsilon_0 \sum_{l,m,n=1}^{3} \chi_{jlmn}^{(3)}(\omega_{TH}, \omega_{FF}, \omega_{FF}, \omega_{FF}) E_{FF,l} E_{FF,m} E_{FF,n}$$

(6)

where $j,l,m,n$ are the Cartesian coordinates, $\varepsilon_0$ is the vacuum electric permittivity and $\chi^{(2)}{}_{jlm}$ and $\chi^{(3)}{}_{jlmn}$ are the instantaneous second and third order susceptibility tensors' components, respectively and $\omega_{TH}$ is the TH angular frequency. In what follows we will consider $\chi_{xxxx}^{(3)} = \chi_{yyyy}^{(3)} = \chi_{zzzz}^{(3)} = \chi_h^{(3)} = \chi_c^{(3)} = 10^{-21}$ m$^2$/V$^2$, values compatible to commercial glasses [47], where the subscript '$h$' and '$c$' stand for host and core, respectively, and assume a relatively low second order susceptibility values, i.e. $\chi_{xxx}^{(2)} = \chi_{yyy}^{(2)} = \chi_{zzz}^{(2)} = \chi_h^{(2)} = \chi_c^{(2)} = 10$ pm/V (all other tensor components are assumed to be zero).

*4.1 Effects of second and third order susceptibilities in the host medium*

We begin our study by first introducing second and third order susceptibilities in the host medium (indicated with $\varepsilon_h$ in Fig. 1). The nonlinear susceptibilities are included only in the 4-*period*-thick region that actually contains the nanoshells and are set to be zero outside. In order to identify the relative nonlinear contribution we neglect all nonlinear terms in the metal described in the previous section. We consider the same probe irradiance ($10 \text{ MW/cm}^2$) and incident wavelength, $\lambda_{FF_2} = 710.6 \text{ nm}$, where $\text{Re}(\varepsilon) = 0$. Our nonlinear calculations (Fig. 6) reveal that both SH and TH have different angular trends when compared with Fig.4 (c). Moreover, we note that the SH signal radiated by the metal (Fig.4 (c)) exceeds the SH generated from the host medium (shown in Fig. 6 (a)) by one order of magnitude, while the THG that arises from the host medium is negligible compared the TH signal that emanates from the metal (Fig.5(c)).

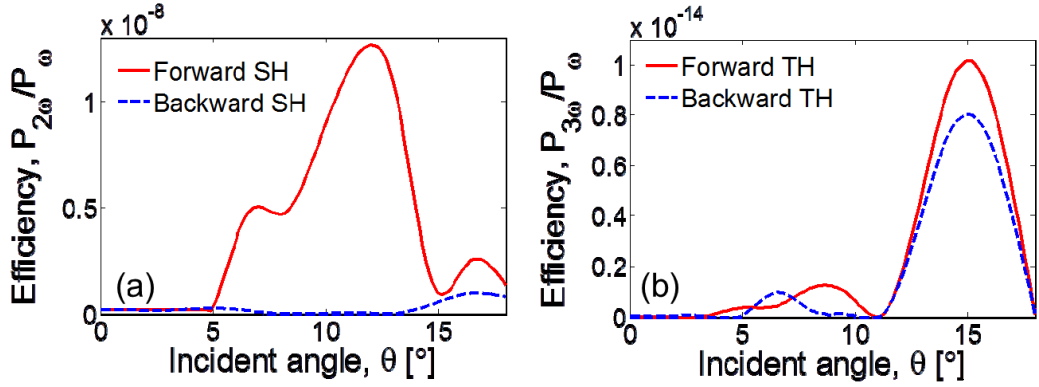

Fig. 6. Forward and backward (a) second and (b) third harmonic conversion efficiencies for the metamaterials described in Sec. 2 at $\lambda_{FF_2} = 710.6 \text{ nm}$ as a function of incident angle. We assume $\chi_h^{(2)} = 10 \text{ pm/V}$ and $\chi_h^{(3)} = 10^{-21} \text{ m}^2/\text{V}^2$ for the host medium. These results have been calculated by means of full-wave (finite element method) simulations.

### 4.2 Effects of second and third order susceptibilities in the nanoshells' core

We repeat the calculations shown in Sec. 4.1 by considering non-zero second and third order susceptibilities in the cores of the nanoshells: $\chi_{xxx}^{(2)} = \chi_{yyy}^{(2)} = \chi_{zzz}^{(2)} = \chi_c^{(2)} = 10 \text{ pm/V}$ and $\chi_{xxxx}^{(3)} = \chi_{yyyy}^{(3)} = \chi_{zzzz}^{(3)} = \chi_c^{(3)} = 10^{-21} \text{ m}^2/\text{V}^2$, probe irradiance is $10 \text{ MW/cm}^2$ and wavelength is $\lambda_{FF_2} = 710.6 \text{ nm}$. Just as before, in this scenario second and third harmonic signals manifest yet a different trend when compared with the same signals arising from metal and from bulk nonlinearities of the host medium external to the nano-shell. In this case SH generation becomes less efficient when only the core of the nanoshells is assumed to be nonlinear (in contrast to the case analyzed in Sec. 4.1 where the host medium is nonlinear). On the other hand, the core-nonlinearity boosts THG by approximately one order of magnitude compared to the case of section 4.1. These contrasting trends in SHG and THG are most likely due to the substantial difference in field localization. For example, at $\lambda_{FF_2} = 710.6$ nm the electric field is well localized inside the gold shell, which makes the SH signal significantly larger when the metal nonlinearity is retained. The same field localization arguments cannot be applied directly to the improvements noted for the TH signal because the third order nonlinearity of the metal is at least four orders of magnitude larger compared to the host or core dielectric's third order nonlinearity , a discrepancy that could account for most of the improvements that we calculate.

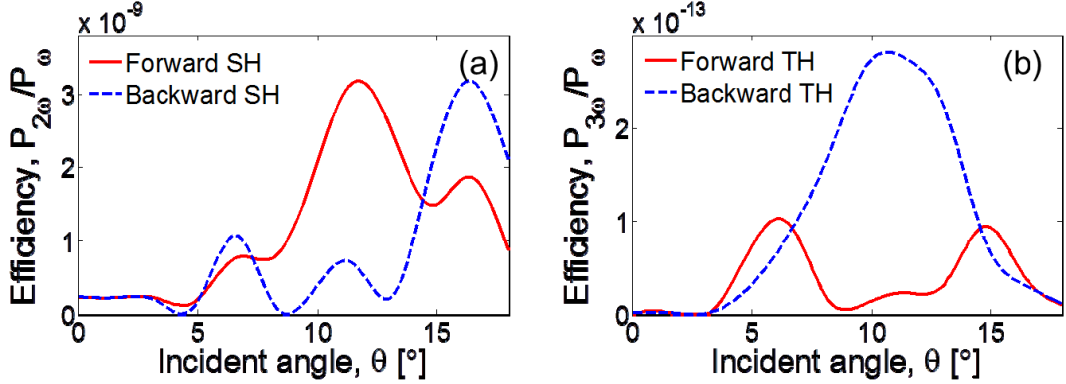

Fig. 7. Forward and backward (a) second and (b) third harmonic conversion efficiencies for the metamaterials described in Sec. 2 at $\lambda_{FF_2} = 710.6$ nm as a function of incident angle. We assumed $\chi_c^{(2)} = 10$ pm/V and $\chi_c^{(3)} = 10^{-21}$ m$^2$/V$^2$ inside the the cores of the nanoshells. These results have been calculated by means of full-wave (finite element method) simulations.

## 5. The role of gain

In order to stress the key role of damping-compensation processes in shaping the linear properties to enhance electric field localization to lower nonlinear thresholds we also calculate transmission, reflection and absorption values for the same structure sketched in Fig.1 by assuming different gain conditions. We first compare the absorption maps obtained with full-wave simulations for nanoshells that contain a silica core doped with Rhodamine 800 with decreasing concentrations: 10 mM (Fig.8 (a)), 7 mM (Fig.8 (b)), 3 mM (Fig.8 (c)) and 0 mM (Fig.8 (d)). Under the same probing conditions the absorption profiles change dramatically while the $\mathrm{Re}(\varepsilon) = 0$ condition red-shifts from $\lambda \approx 710.6$ nm at 10 mM to $\lambda \approx 714.3$ nm at 7 mM, and from to $\lambda \approx 715.9$ nm at 3 mM to $\lambda \approx 717.7$ nm for 0 mM. We stress that the $\mathrm{Re}(\varepsilon) = 0$ conditions for different concentrations are determined by means of the Nicolson-Ross-Weir method applied to a metamaterial slab of finite thickness under normal incidence conditions [26-27]. By comparing the results in Fig. 8 one may clearly infer that the majority of the features that characterize absorption (and implicitly also reflection and transmission – not shown here) arise from the ability to compensate damping in the metamaterial. Although a predominant, wideband absorption feature is present even without damping compensation (Fig. 8(d)), this spectral/angular feature narrows and bends as dye concentration is increased and damping is compensated. On the other hand, while the effective damping of the structure is being compensated ( $\mathrm{Im}(\varepsilon)$ goes from $\approx 0.4$ for 0 mM to $\approx 10^{-4}$ for 10 mM, at their respective zero-crossing point for $\mathrm{Re}(\varepsilon)$ ), absorption is still dramatically high because field localization is boosted. This conclusion is straightforward if one considers the absorption is proportional to the product $\mathrm{Im}(\varepsilon_{\mathrm{eff}})\,|\,\mathbf{E}\,|^{2}$. High concentrations of dye molecules may impact overall damping compensation due to the presence of fluorescence quenching and other non-radiative phenomena [51]. The quenching effect leads to a reduction of gain in the system, although an analytical treatment of the molecule-nanoparticle interaction is probably needed to estimate its real impact. Alternative designs that for example use quantum dots [8] may be employed here.

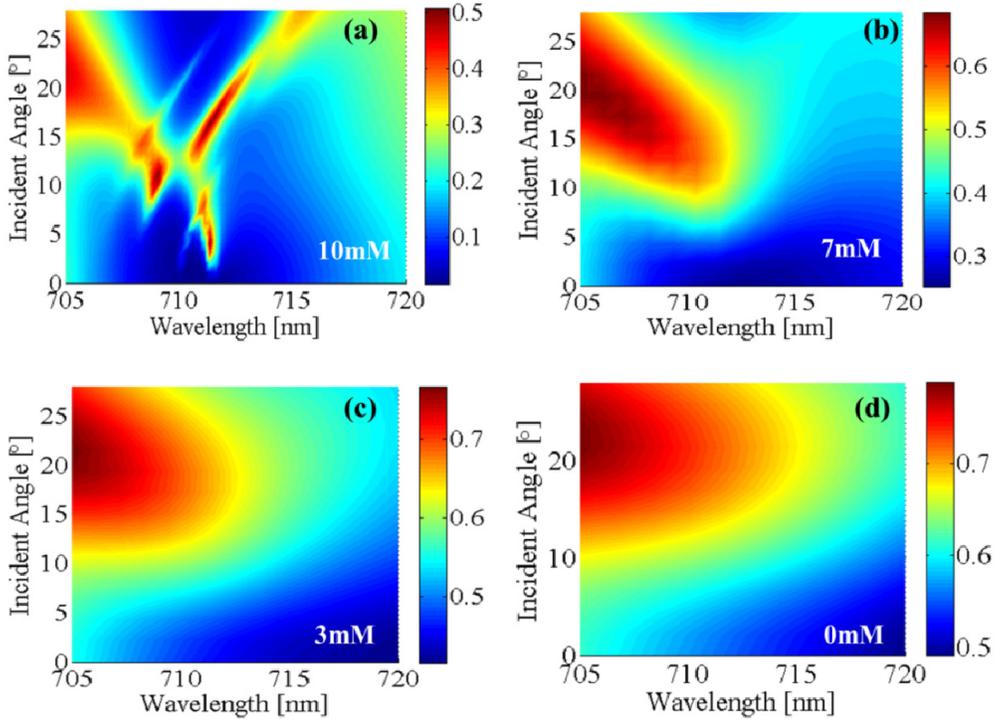

Fig. 8. Full-wave calculations of absorption as a function of wavelength and incident angle for (a) 10 mM, (b) 7 mM, (c) 3 mM and (d) 0 mM of Rhodamine 800 dispersed in the silica cores of the nanoshells illustrated in Fig. 1. These results have been calculated by means of full-wave (finite element method) simulations.

An immediate consequence of the differences in field localization among the four structures with variable damping-compensation is a variation in the strength of harmonic signal. In order to understand how critical is the introduction of proper gain (or alternatively, proper pumping condition) to the nonlinear response we calculate forward and backward TH signals (arising from the metal shells only) from the four structures with absorption profiles in Fig. 8 where $Re(\varepsilon) = 0$: $\lambda \approx 710.6$ nm for 10 mM (Fig. 9(a)), $\lambda \approx 714.3$ nm for 7 mM (Fig. 9(b)), $\lambda \approx 715.9$ nm for 3 mM (Fig. 9(c)) and $\lambda \approx 717.7$ nm for 0 mM (Fig. 9(d)). The detrimental effect induced by the reduction or complete elimination of the active material from the cores of the nanoshells is quite evident from an inspection of Fig. 9. A dramatic drop in TH efficiencies by about three orders of magnitude is observed when gain is eliminated. These results should also be compared with THG from a 5 nm-thick gold film, which exhibits a very similar angular behavior and conversion efficiency (Fig. 5(a)). The results shown in Fig. 9 demonstrate quite clearly how the introduction of gain in the metamaterial not only gives a different angular trend to the generated signals but also improves significantly conversion efficiencies between two and three orders of magnitude. Similar results are also expected for second harmonic generation.

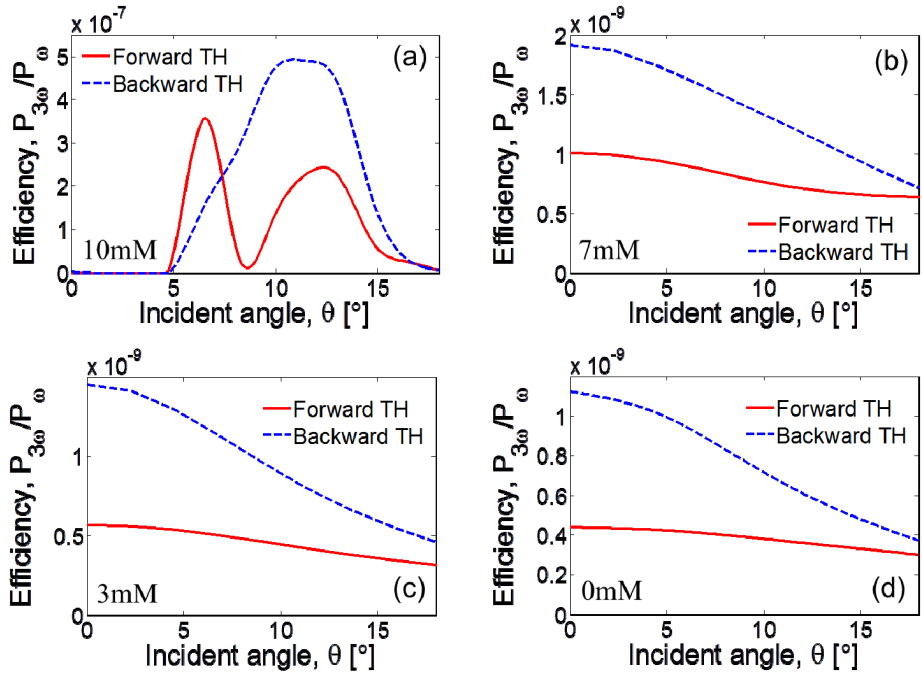

Fig. 9. Forward and backward third harmonic conversion efficiency $\eta_{TH} = P_{3\omega} / P_{\omega}$ for the metamaterial described in Sec. 2 with (a) 10 mM, (b) 7 mM, (c) 3 mM and (d) 0 mM of Rodhamine 800 dispersed in the cores of the nanoshells. The simulations are performed with a probe tuned at (a) $\lambda = 710.6$ nm, (b) $\lambda = 714.3$ nm, (c) $\lambda = 715.9$ nm and (d) $\lambda = 717.7$ nm, where the metamaterials effective perrmittivity show $Re(\varepsilon) = 0$. These results have been calculated by means of full-wave (finite element method) simulations.

## 6. Conclusions

We have studied second and third harmonic generation from metamaterials composed of plasmonic nanoshells working in a damping-compensation regime, to stress the ability of engineered materials exhibiting NZP properties to boost nonlinear processes. The introduction of a gain material into the cores of the nanoshells not only results in peculiar, narrow transmission, reflection and absorption features but also shapes dramatically the nonlinear response increasing both second and third harmonic efficiencies, which may be evaluated on a case-by-case basis for each resonant feature of the metamaterial. These results demonstrate how efficient, singularity-driven, low-threshold second and third harmonic generation processes may originate from the strong enhancement of the electric field achievable in metamaterials with NZP properties. Moreover, the investigation of both linear and nonlinear regimes performed by reducing the concentration of gain material or pump intensity proves the extreme sensibility of this regime to pump conditions and serves as a guide for future experimental works within the framework of damping compensation.